\documentclass[preprint,aps,superscriptaddress,nofootinbib,tightenlines]{revtex4}

\usepackage{epsfig}

\def\OMIT#1{{}}

\def\si{^1 \hskip -0.03in S _0}
\def\siii{^3 \hskip -0.025in S _1}
\def\diii{^3 \hskip -0.03in D _1}
\def\pislash{ {\pi\hskip-0.54em /} }

\def\nopi{ {\rm EFT}(\pislash) }
\def\nopit{ {\rm dEFT}(\pislash) }

\def\tr{di-baryon}
\def\trs{di-baryons}

\begin{document}


\preprint{\vbox{
\hbox{UNH-04-07}
\hbox{NT@UW-04-027}
}}

\phantom{ijk}
\vskip 0.5cm
\title{DVCS-Dissociation of the Deuteron and the EMC Effect}

\author{\bf Silas R.~Beane}
\affiliation{Department of Physics, University of New Hampshire,
Durham, NH 03824-3568.}
\affiliation{Jefferson Laboratory, 12000 Jefferson Avenue, 
Newport News, VA 23606.}
\author{\bf Martin J.~Savage}
\affiliation{Department of Physics, University of Washington, 
Seattle, WA 98195-1560.}
\affiliation{Center for Theoretical Physics, MIT,
Cambridge, MA 02139.\\
\qquad
}

\vphantom{}
\vskip 0.5cm
\begin{abstract} 
\noindent 
The break-up of the deuteron during deeply-virtual Compton scattering,
$\gamma^* d\rightarrow \gamma^{(*)} n p $, is explored.  In the
effective field theory describing nucleon dynamics at momenta below
the pion mass, the EMC effect results from four-nucleon interactions
with the twist-2 operators, appropriate for describing forward, and
near-forward, matrix elements in the two-nucleon system.  We point out
that the break-up of the deuteron to low-energy final states during
deeply-virtual Compton scattering is a process with which to explore
strong-interaction physics closely related to that responsible for the
EMC effect.  The single-nucleon contribution to the break-up depends
on the moments of the spin-dependent structure functions and
contributions from local four-nucleon operators.  Experimental
deviations from the single-nucleon prediction would provide a
probe of strong interactions complimentary to the EMC effect.
\end{abstract}

\maketitle

\noindent A naive glance at QCD would suggest that the nucleus is a
non-perturbative collection of quarks and gluons with no discernible
structure.  However, this is not the case, and a remarkable aspect of
nuclear physics is that nuclei may be viewed as a collection of
relatively weakly-interacting nucleons.  The small binding energy of
the deuteron implies that it very-nearly resembles a free proton and
neutron; its magnetic moment is very close to the sum of proton and
neutron magnetic moments, and its structure functions are very close
to the sum of the nucleon structure functions.  Deviations from the
single-nucleon structure functions, known as the EMC
effect~\footnote{The expression ``EMC effect'' usually refers to
deviations from single-nucleon structure in nuclei with $A\gg 2$. Here
we apply the expression to the deuteron and therefore, in the present
context, the EMC effect is relevant to extractions of the neutron
structure functions.}, were experimentally
observed~\cite{Aubert:1983xm} in nuclei some twenty years ago and
caused a great deal of excitement,
e.g.~\cite{Arneodo:1992wf,Geesaman:1995yd,Piller:1999wx,Norton:2003cb}.  However, as
the nucleons do interact, albeit weakly, one expects matrix elements
of all electro-weak probes to be modified from their single-nucleon
values, and the EMC effect should be considered to be nothing more
than another demonstration of this.  Significant efforts have gone
into quantitatively explaining the EMC effect in explicit models of
nuclei, e.g. Ref.~\cite{Smith:2003hu,Steffens:2004yb}.  One would like
to uncover the dominant strong-interaction mechanism that allows for a
quantitative explanation of the effect, but given that one cannot
accurately predict the deuteron quadrupole moment from modern
potential models~\footnote{ The quadrupole moment of the deuteron is
successfully described in effective field theory by the inclusion of a
local operator at next-to-leading order in the expansion~\cite{Chen:1999tn}.  The
discrepancy between potential model calculations and the experimental
value is due to the omission of physics at distance scales smaller
than the Compton wavelength of the pion, which may or may not be
describable by a meson-exchange mechanism.}  to better than $\sim
5\%$, an approximate explanation would suffice at present.

Within the framework of the pionless effective field theory, $\nopi$,
it is easy to understand and describe the EMC effect.  By constructing
all the operators required to reproduce the matrix elements of the
twist-2 operators in multi-nucleon systems, one sees that operators
involving more than one nucleon are not forbidden by the symmetries of
the strong interaction, and therefore must be present.  While
observation of the EMC effect twenty years ago may have been
surprising to some, in fact, its absence would have been far more
surprising.  In this work we write down the most general set of
operators describing the interaction between two nucleons in an S-wave
that reproduce the matrix elements of the twist-2 operators.  As the
origin of the EMC effect is strong-interaction physics at distance
scales of order the Compton wavelength of the pion or smaller, there
will be an analogous deviation from the single-nucleon prediction for
transitions between two nucleons in the $\siii$ channel and the $\si$
channel.  Therefore, we suggest that an experimental determination of
the break-up of the deuteron during Deeply-Virtual Compton Scattering
(DVCS), with very low-energy final-state nucleons, provides a probe of
a purely-nuclear effect which is complimentary to the EMC effect.

In any given hadron, the moments of the parton distribution functions are given by 
\begin{eqnarray}
\left\langle x^{n-1}\right\rangle _{ q_V} &=&\int_{0}^{1}dx\,x^{n-1}\left(\,
q\left( x\right)\, +\,\left( -1\right) ^{n}\overline{q}\left(x \right)\, \right)  \ ; 
     \nonumber \\
\left\langle x^{n-1}\right\rangle _{ q_A} &=&\int_{0}^{1}dx\,x^{n-1}\left(\,
\Delta q\left( x\right)\, +\,\left( -1\right) ^{n-1}\Delta \overline{q%
}\left( x\right) \,\right) \ ,
\end{eqnarray}
where $q$ ($\overline{q}$) is the quark (antiquark) spin-averaged
distribution and $\Delta q$ $(\Delta \overline{q})$ is the helicity
distribution (we do not consider the transversity distribution).  From
QCD, these moments have the clear interpretation as the forward matrix
elements of local operators,
\begin{eqnarray}
\theta_{V,\ \mu_1 ..\mu_n }^{(n),0} & = &  (i)^{n-1}\ 
\overline{q} \ \gamma_{\{ \mu_1}\ 
\stackrel{\leftrightarrow}{D}_{\mu_2}\ 
...\ 
\stackrel{\leftrightarrow}{D}_{\mu_n\} }\ q
\nonumber\\
\theta_{V,\ \mu_1 ..\mu_n }^{(n),b} & = & (i)^{n-1}\ 
\overline{q} \ \gamma_{\{ \mu_1}\ 
\stackrel{\leftrightarrow}{D}_{\mu_2}\ 
...\ 
\stackrel{\leftrightarrow}{D}_{\mu_n\} }\ \tau^b\ q
\nonumber\\
\theta_{A,\ \mu_1 ..\mu_n }^{(n),0} & = & (i)^{n-1}\ 
\overline{q} \ \gamma_{\{ \mu_1}\ 
\stackrel{\leftrightarrow}{D}_{\mu_2}\ 
...\ 
\stackrel{\leftrightarrow}{D}_{\mu_n\} }\ \gamma_5\ q
\nonumber\\
\theta_{A,\ \mu_1 ..\mu_n }^{(n),b} & = & (i)^{n-1}\ 
\overline{q} \ \gamma_{\{ \mu_1}\ 
\stackrel{\leftrightarrow}{D}_{\mu_2}\ 
...\ 
\stackrel{\leftrightarrow}{D}_{\mu_n\} }\ \gamma_5\ \tau^b\ q
\ \ \ ,
\label{eq:twisttwoops}
\end{eqnarray}
where the indices enclosed by $\{ ... \}$ are symmetrized and their
traces are removed.  An analysis of matrix elements of these operators
in the single-nucleon sector has been carried out at the one-loop
level in the chiral
expansion~\cite{Arndt:2001ye,Chen:2001eg,Chen:2001et,Chen:2001nb,Chen:2001pv}.
Those works have been extended from QCD to quenched
QCD~\cite{Chen:2001gr} and partially-quenched
QCD~\cite{Chen:2001yi,Beane:2002vq} for applications to lattice QCD
simulations.  In addition to forward matrix elements of these
operators, off-forward matrix elements have been considered~\cite{Belitsky:2002jp}, and the
issue of soft-pion production during DVCS has been 
explored~\cite{Chen:2003jm,Guichon:2003ah}~\footnote{
The errors made in Ref.~\cite{Guichon:2003ah} and addressed in
Ref.~\cite{Chen:2003jm} have been replicated
in Ref.~\cite{Kivel:2004bb}. This latter paper
does not properly account for the fact that chiral symmetry is
spontaneously broken, and thus leads to results ---like $g_A=1$--- which,
while possibly interesting, are not relevant to QCD.}.
For very low-energy processes, those involving momentum and energy
transfers much less than the pion mass, one can work in $\nopi$.  In
this energy regime, the matrix elements of the operators in
eq.~(\ref{eq:twisttwoops}) between single-nucleon states are
reproduced by
\begin{eqnarray}
\theta_{V,\ \mu_1 ..\mu_n }^{(n),0} & \rightarrow & 
M^{n-1}\ 
\langle x^{n-1}\rangle_{ q_V^{(0)}}\ \ 
v_{\{ \mu_1} v_{\mu_2} \ ...\ v_{\mu_n\} } \ \ 
\overline{N} N
\ +\ ...
\nonumber\\
\theta_{V,\ \mu_1 ..\mu_n }^{(n),b} & \rightarrow & 
M^{n-1}\ 
\langle x^{n-1}\rangle_{ q_V^{(1)}}\ \ 
v_{\{ \mu_1} v_{\mu_2} \ ...\ v_{\mu_n\} } \ \ 
\overline{N} \ \tau^b\ N
\ +\ ...
\nonumber\\
\theta_{A,\ \mu_1 ..\mu_n }^{(n),0} & \rightarrow & 
M^{n-1}\ 
\langle x^{n-1}\rangle_{ q_A^{(0)}}\ \ 
\ v_{ \{ \mu_1} v_{\mu_2} \ ...\ v_{\mu_{n-1}} \ \ 
\overline{N} \ S_{\mu_n \} }\ N
\ +\ ...
\nonumber\\
\theta_{A,\ \mu_1 ..\mu_n }^{(n),b} & \rightarrow & 
M^{n-1}\ 
\langle x^{n-1}\rangle_{ q_A^{(1)}}\ \ 
v_{ \{ \mu_1} v_{\mu_2} \ ...\ v_{\mu_{n-1}} \ \ 
\overline{N}\  S_{\mu_n \} }\  \tau^b\ N
\ +\ ...
\ \ \ ,
\label{eq:isoscalarvector}
\end{eqnarray}
where $\langle x^{p}\rangle_{ q_V^{(0)}}$ and $\langle x^{p}\rangle_{
q_A^{(0)}}$ are the $p$'th moments of the isoscalar vector and
axial-vector PDF's in the nucleon, respectively, and $\langle
x^{p}\rangle_{ q_V^{(1)}}$ and $\langle x^{p}\rangle_{ q_A^{(1)}}$ are
the $p$'th moments of the isovector PDF's in the nucleon,
respectively.  $M$ is the nucleon mass. 
For $n=1$  baryon-number conservation gives 
$\langle 1\rangle_{ q_V^{(0)}}=3$; isospin-conservation gives
$\langle 1\rangle_{ q_V^{(1)}}=1$; 
and matching to the axial-vector currents gives 
$\langle 1\rangle_{ q_A^{(0)}}=g_0$, the singlet axial charge~\footnote{
The ``$q$'' subscripts on the operator coefficients 
in eq.~(\ref{eq:isoscalarvector}) and eq.~(\ref{eq:twobod})
indicate that
we are considering matrix elements of the quark operators only,
and not the purely-gluonic operators 
--including the anomaly-- which mix with the quark operators under
renormalization. 
The singlet axial charge defined here, $g_0$, is therefore scale dependent.}, 
and 
$\langle 1\rangle_{ q_A^{(1)}}=g_A$, the isovector axial charge. 
The nucleon four-velocity is
$v$, and in the nucleon rest frame becomes $v=(1,0,0,0)$.  The
ellipses denote terms that are higher order in the derivative
expansion, suppressed by powers of $p/m_\pi$.

When considering the parton distributions of the deuteron, the
simplest nucleus, insertions of the operators of
eq.~(\ref{eq:isoscalarvector}) defines the impulse approximation.
Given the operator structure, it is clear that one simply recovers the
naive sum of contributions from the proton and neutron.  However, it
is also clear that, in the effective field theory
expansion, there are contributions from
higher-dimensional operators involving more 
nucleons.  The leading-order interactions involving
two-nucleons are
\begin{eqnarray}
\theta_{V,\ \mu_1 ..\mu_n }^{(n),0} & \rightarrow & 
M^{n-1}\ 
\left[\ 
\alpha^{(n)}_{q}\ v_{\{ \mu_1} v_{\mu_2} \ ...\ v_{\mu_n\} } \ 
\left(N^T P^a_1 N\right)^\dagger 
N^T P^a_1 N 
\right.
\nonumber\\
& & 
\left. \ + \ 
\beta^{(n)}_{0; q}\ v_{\{ \mu_1} v_{\mu_2} \ ...\ v_{\mu_n\} } \ 
\left(N^T P^\alpha_3 N\right)^\dagger 
N^T P^\alpha_3 N 
\right.
\nonumber\\
& & 
\left.
\ + \ 
\beta^{(n)}_{2; q}\ v_{\{ \mu_3} v_{\mu_4} \ ...\ v_{\mu_n} \ 
\left(N^T P^{\mu_1}_3 N\right)^\dagger 
N^T P^{\mu_2\} }_3 N 
\ \right]
\nonumber\\
\theta_{V,\ \mu_1 ..\mu_n }^{(n),a} & \rightarrow & 
M^{n-1}\ \gamma^{(n)}_{q}\ 
i \varepsilon^{abc}\ 
v_{\{ \mu_1} v_{\mu_2} \ ...\ v_{\mu_n \} } \ 
\left(N^T P^b_1 N\right)^\dagger 
N^T P^c_1 N 
\nonumber\\
\theta_{A,\ \mu_1 ..\mu_n }^{(n),0} & \rightarrow & 
M^{n-1}\ \rho^{(n)}_{q}\ 
\ v_{\{ \mu_1} v_{\mu_2} \ ...\ v_{\mu_{n-1}} \ 
i \varepsilon^{\mu_n\}\alpha\beta\gamma}\ v_\alpha
\left(N^T P^{\beta}_3 N\right)^\dagger 
N^T P^{\gamma}_3 N 
\nonumber\\
\theta_{A,\ \mu_1 ..\mu_n }^{(n),a} & \rightarrow & 
M^{n-1}\ 
\sigma^{(n)}_{q}\ v_{\{ \mu_1} v_{\mu_2} \ ...\ v_{\mu_{n-1}} \ 
\left(N^T P^a_1 N\right)^\dagger 
N^T P^{\mu_n\} }_3 N 
\ +\ {\rm h.c.}
\ \ \ ,
\label{eq:twobod}
\end{eqnarray}
where with $n=1$ we have $\beta^{(1)}_{2;q}=0$ (due to available
Lorentz indices) and baryon-number conservation gives
$\alpha^{(1)}_q=\beta^{(1)}_{0;q}=0$; isospin-conservation gives
$\gamma^{(1)}_q=0$; and matching to the axial-vector current gives
$\rho^{(1)}_q=-2L_{2,A}$ and
$\sigma^{(1)}_q=L_{1,A}$~\cite{Butler:1999sv}~\footnote{The
cross-sections for the weak disintegration of the deuteron and other
two-nucleon weak processes depend on the parameter $L_{1,A}$. The
axial operator with coefficient $L_{2,A}$ contributes ---together with the sum of the
neutron and proton contributions--- to the
strangeness content of the deuteron.}.
The objects $P^a_1$ and
$P^\mu_3$~\cite{Kaplan:1998tg,Kaplan:1998we,Kaplan:1998sz} are the
spin-isospin projectors for the $\si$ and $\siii$ channels,
respectively,
\begin{eqnarray}
P_1^a & = & {1\over\sqrt{2}}\ \tau_2\tau^a \otimes S_2
\ \ ,\ \ 
P_3^\mu \ = \ {\sqrt{2}}\  \tau_2 \otimes S_2\, S^\mu
\ \ \ ,
\end{eqnarray}
where $S^\mu$ is the usual spin operator which acts on the nucleon fields.
The contributions of such operators to the parton distributions of the
deuteron can be computed straightforwardly in 
$\nopi$~\cite{Chen:1999tn,vanKolck:1997ut}~\footnote{
The effective field 
theory can be constructed in terms of operators that do not vanish on the particle mass-shells.
This is true for any consistent theory, and operators that only contribute off mass-shell can always be 
eliminated.
}.  
However, it
is convenient to switch notation to that of $\nopi$ constructed in
terms of di-baryon fields~\cite{Kaplan:1996nv,Beane:2000fi}, $\nopit$,
in which the leading effects of the finite effective range are
resummed.

In this work we will discuss relevant aspects
of the \tr\ formalism, $\nopit$~\cite{Kaplan:1996nv,Beane:2000fi,Ando:2004mm}.  In terms of
nucleon and \tr\ degrees of freedom, the leading-order low-energy
strong interactions for $|{\bf p}|\ll m_\pi/2$ are described by a
Lagrange density of the form
\begin{eqnarray}
{\cal L} & = & 
N^\dagger\left[\ i\partial_0 + {\nabla^2\over 2 M}\right] N
\ +\ t_\mu^\dagger\left[\ i\partial_0 + {\nabla^2\over 4 M}-\Delta_3\right] t^\mu
\ -\ s_a^\dagger\left[\ i\partial_0 + {\nabla^2\over 4 M}-\Delta_1\right] s^a
\nonumber\\
& &
+ y_3 \left[\ t_\mu^\dagger\  N^T P_3^\mu N\ +\ {\rm h.c.}\right]
- y_1 \left[\ s_a^\dagger\  N^T P_1^a N\ +\ {\rm h.c.}\right]
\ \ \ ,
\label{eq:lagST}
\end{eqnarray}
where $N$ is the nucleon annihilation operator, 
we have used the four-vector $t_\mu=(0,t_j)$ with $t^j$ the $\siii$
\tr\ annihilation operator with spin-index $j$, and $s^a$ is the $\si$
\tr\ annihilation operator with isospin-index $a$.  In the Lagrange
density in eq.~(\ref{eq:lagST}), a factor of the nucleon mass $M$ ($2M$) has
been absorbed into the definition of the nucleon fields (\tr\ fields).
The S-wave interactions are enhanced by a factor of the expansion parameter, $1/Q$, and are
treated non-perturbatively. However the interactions that induce
mixing with higher partial waves, e.g. $\siii-\diii$ mixing, are
suppressed by at least $Q^2$ and so only S-wave to S-wave interactions
are required to the order we are working.  To recover the scattering
amplitudes in both S-wave channels, the constants that appear in
eq.~(\ref{eq:lagST}) are found to be~\cite{Kaplan:1996nv}
\begin{eqnarray}
y_3^2 & = & {8\pi\over M^2 r_3}
\ \ ,\ \ 
y_1^2\ =\ {8\pi\over M^2 r_1}
\ \ ,\ \ 
\Delta_3 \ = \ {2\over M r_3} \left( {1\over a_3}-\mu\ \right)
\ \ ,\ \ 
\Delta_1 \ = \ {2\over M r_1} \left( {1\over a_1}-\mu\ \right)
\ ,
\label{eq:tdefs}
\end{eqnarray}
where $\mu$ is the renormalization scale, $a_3$ and $r_3$ are the
scattering length and effective range in the $\siii$ channel, and
$a_1$ and $r_1$ are the scattering length and effective range in the
$\si$ channel.  Insertions of the twist-2 operators give rise to
operators of the form
\begin{eqnarray}
\theta_{V,\ \mu_1 ..\mu_n }^{(n),0} & \rightarrow & 
M^{n-1}\ \left[\ 
\tilde \alpha^{(n)}_{q}\ v_{\{ \mu_1} v_{\mu_2} \ ...\ v_{\mu_n \} } \ 
s^\dagger_a s^a 
\right.
\nonumber\\
& & 
\left. \qquad\qquad \ + \ 
\tilde \beta^{(n)}_{0; q}\ v_{\{ \mu_1} v_{\mu_2} \ ...\ v_{\mu_n \} } \ 
t^\dagger_\alpha t^\alpha
\ + \ 
\tilde \beta^{(n)}_{2; q}\ v_{\{ \mu_3} v_{\mu_4} \ ...\ v_{\mu_n} \ 
t^\dagger_{\mu_1} t^{\mu_2 \} }
\ \right]
\nonumber\\
\theta_{V,\ \mu_1 ..\mu_n }^{(n),3} & \rightarrow & 
M^{n-1}\ 
\tilde\gamma^{(n)}_{q}\ i \varepsilon^{3bc}
\ v_{\{ \mu_1} v_{\mu_2} \ ...\ v_{\mu_n \} } \ 
s_b^\dagger\ s_c
\nonumber\\
\theta_{A,\ \mu_1 ..\mu_n }^{(n),0} & \rightarrow & 
M^{n-1}\ \tilde\rho^{(n)}_{q}\ 
\ v_{\{ \mu_1} v_{\mu_2} \ ...\ v_{\mu_{n-1}} \ 
i \varepsilon^{\mu_n\}\alpha\beta\gamma}\ v_\alpha\ 
t_\beta^\dagger\  t_\gamma
\nonumber\\
\theta_{A,\ \mu_1 ..\mu_n }^{(n),3} & \rightarrow & 
M^{n-1}\ 
\tilde \sigma^{(n)}_{q}\ v_{\{ \mu_1} v_{\mu_2} \ ...\ v_{\mu_{n-1}} \ 
\left[\ s^\dagger_3 \ t^{\mu_n \} }\ + \ {\rm h.c.}\ \right] \ \ .
\label{eq:twobodt}
\end{eqnarray}
With $n=1$ we have $\tilde\beta^{(1)}_{2;q}=0$ and baryon-number
conservation gives $\tilde\alpha^{(1)}_q=-\tilde\beta^{(1)}_{0;q}=6$;
isospin-conservation gives $\tilde\gamma^{(1)}_q=2$; and matching to
the axial-vector current gives
$\tilde\rho^{(1)}_q=-2\ell_{2,A}/2M\sqrt{r_1r_3}$ and
$\tilde\sigma^{(1)}_q=\ell_{1,A}/2M\sqrt{r_1r_3}$~\cite{Detmold:2004qn}.
(Recent work has shown that it is possible to compute the coefficients
of the operators in eq.~(\ref{eq:twobodt}) in lattice QCD by making
use of background-field methods~\cite{Detmold:2004qn,Detmold:2004kw}.)

\begin{figure}[!t]
\centerline{{\epsfxsize=4.in \epsfbox{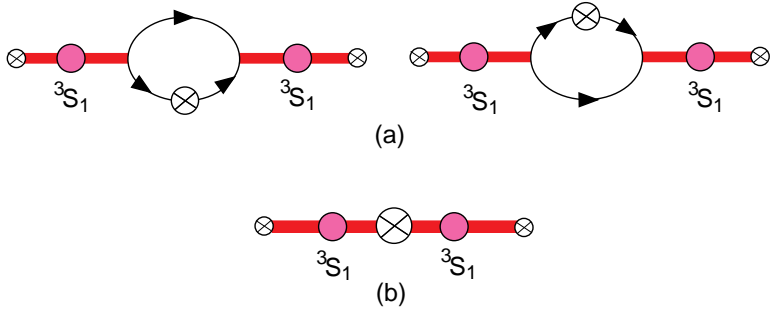}}} 
\vskip 0.15in
\noindent
\caption{\it 
Diagrams in $\nopit$ that provide the leading-order contribution to the 
matrix elements of the twist-2 operators in the deuteron.
The thick solid lines with solid circles at their midpoint correspond to fully-dressed
\trs, while the thin lines correspond to nucleons.  
The crossed circles correspond to insertions of the twist-2 operators.
}
\label{fig:emc}
\vskip .2in
\end{figure}
It is now straightforward to determine the matrix elements of the
relevant operators between deuteron states.  Keeping only the
$\Delta L=0$ contribution from the diagrams shown in
Fig.~\ref{fig:emc}, the matrix elements of the isoscalar-vector quark
operators at momentum transfer ${\bf q}\ll m_\pi$ are
\begin{eqnarray}
\langle d | \ \theta_{V,\ \mu_1 ..\mu_n }^{(n),0}\  | d \rangle
& = & 
M^{n-1}\ 
v_{\{ \mu_1} v_{\mu_2} \ ...\ v_{\mu_n \} } \ \ 
\left[\ 
2\ \langle x^{n-1}\rangle_{ q_V^{(0)}}\ F_C\left( |{\bf q}|\right) 
\ -\ \tilde\eta_{0; q}^{(n)} {\gamma r_3\over 1- \gamma r_3}
\ \right]
\ \ \ ,
\label{eq:emc}
\end{eqnarray}
where the expression is given in terms of the electric charge form factor
 \begin{eqnarray}
F_C ( |{\bf q}| )
& = & 
{\gamma r_3\over 1-\gamma r_3}
\left[{4\over |{\bf q}|\  r_3}
\tan^{-1}\left({ |{\bf q}|\over 4\gamma}\right)\ -\ 1\ \right]\ \ \ .
\label{eq:CFFtot}
\end{eqnarray}
The single-nucleon contribution has been  made explicit by writing
\begin{eqnarray}
-\tilde\beta^{(n)}_{0; q} & = & 2 \langle x^{n-1}\rangle_{ q_V^{(0)}}
\ +\ \tilde \eta_{0; q}^{(n)}
\ \ \ .
\label{eq:etadef}
\end{eqnarray}
Here we see explicitly that the contribution from the nucleon structure functions
(first term in the square brackets of eq.~(\ref{eq:emc})) is accompanied by
a nuclear contribution (second term in the square brackets of eq.~(\ref{eq:emc})),
making clear the origin of the EMC effect. Notice that for the
baryon-number current ($n=1$), $\tilde \eta_{0; q}^{(1)}=0$ and the
matrix element of eq.~(\ref{eq:emc}) is entirely given by the 
leading-order electric
charge form factor of the deuteron; as emphasized in
Ref.~\cite{Ando:2004mm}, it is a peculiar feature of $\nopit$ (unlike
$\nopi$) that higher-dimensional local \tr\ operators contain
contributions that should be interpreted as leading-order in the
effective field theory expansion.
Several points are worth emphasizing as they bear on conventional
explanations of the EMC effect: 
1) relativistic corrections appear at the next order in the effective
field theory expansion and can be systematically included. These
corrections have been computed for various deuteron properties in $\nopi$
and found to be small~\cite{Chen:1999tn}; for static processes these
corrections scale as $\gamma^2/M^2$. Notice that this expansion parameter
is suppressed relative to the expansion parameter
$\gamma/m_\pi$ characteristic of the size of the purely nuclear effects in
$\nopi$ and $\nopit$. 
2) ``Fermi motion''  (or more appropriately,  ``zero-point energy'') of the deuteron,
is included in $\nopi$ and $\nopit$; terms of the form ${\vec p}^{\;2}/2m$ where
${\vec p}$ is a nucleon momentum are summed to all orders through the 
non-relativistic nucleon propagators from eq.~(\ref{eq:lagST}).
3) As pointed out above, in $\nopi$ and $\nopit$ D-state admixtures enter at higher order in the effective field
theory expansion~\cite{Chen:1999tn}.

There is also a ``tensor EMC effect''.  This is induced by the
operator in eq.~(\ref{eq:twobodt}) with coefficient
$\tilde\beta_{2;q}^{(n)}$, and corresponds to transferring $\Delta
L=2$ to the hadronic system.  This does not contribute to DIS, as it
is not a forward matrix element, but will contribute to DVCS.  If the
deuteron polarization in the final state could be measured after DVCS
on a polarized deuteron target, then in principle this object could be
measured.  Experimentally, such a measurement will not be easy.

A different situation in which the effects of the interactions between
nucleons during DVCS will lead to modifications from the impulse
approximation is the break-up of the deuteron~\footnote{
For a recent discussion of the process $e+d\rightarrow e+n+p$
in the $1.5\leq Q^2<4~{\rm GeV}^2$ range,
see Ref.~\cite{Sargsian:2002wc}.
Deviations from the impulse approximation in DVCS on nuclei has been considered
recently in Refs.~\cite{Kirchner:2003wt} and \cite{Freund:2003ix}.
}.
Such deviations are, in some sense, intimately related to the EMC
effect as they result from different four-nucleon operators connecting
S-wave states.  They are analogous to the local four-nucleon operators
which cause deviations from the impulse approximation for low-energy
$np\rightarrow d\gamma$ (which result from interactions between
nucleons that are not related by gauge-invariance to the nucleon-nucleon scattering
amplitude). It may or may not be the case that pion-exchange currents
can reproduce much of this contribution.
\begin{figure}[!t]
\centerline{{\epsfxsize=4.in \epsfbox{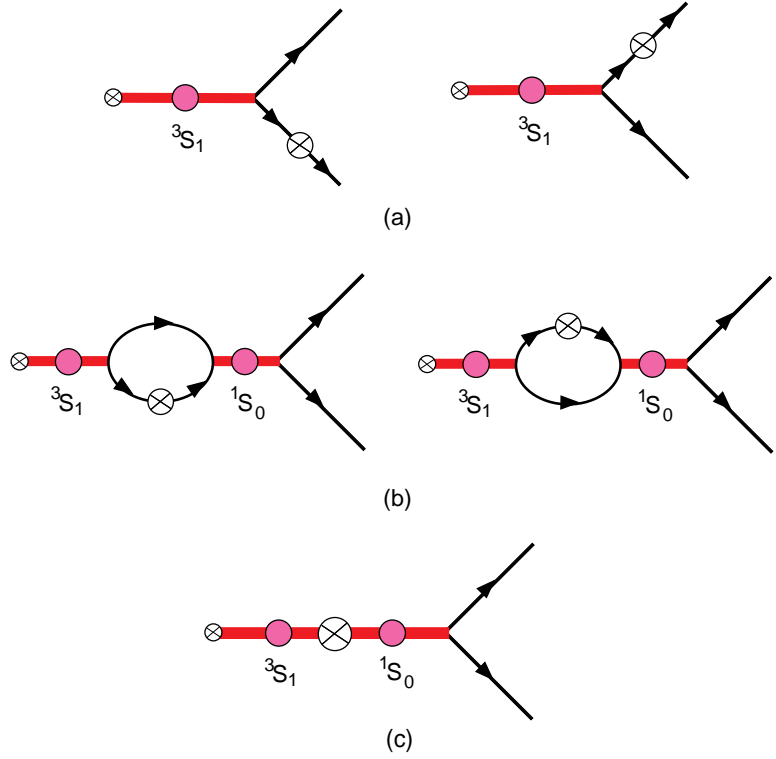}}} 
\vskip 0.15in
\noindent
\caption{\it 
  Diagrams in $\nopit$ that provide the leading-order contribution to the 
matrix elements for the break-up of  the deuteron
during DVCS.
The thick solid lines with solid circles at their midpoint correspond to fully-dressed
\trs, while the thin lines correspond to nucleons.  
The crossed circles correspond to insertions of the twist-2 operators.
}
\label{fig:disso}
\vskip .2in
\end{figure}

It is straightforward to show that 
for very-low energy and momentum transfers to the hadronic system,
the amplitude for DVCS-dissociation
of the deuteron can be written as
\begin{eqnarray}
{\cal A} & = & X\ 
M^{n-1}\ 
v_{ \{ \mu_1} v_{\mu_2} \ ...\ v_{\mu_{n-1}} \ 
\varepsilon_{\mu_n \} }^{(d)}\ 
U_n^T\sigma_2 U_p
\ \ \ ,
\label{eq:dvcsamp}
\end{eqnarray}
for the neutron and proton in a $\si$ final-state, where
$\varepsilon_{\mu}^{(d)}$ is the deuteron polarization vector, and
$U_{n,p}$ are the spinors associated with the neutron and proton
respectively.  For the sake of simplicity, we give an expression for
the coefficient $X$ in eq.(\ref{eq:dvcsamp}), resulting from the
diagrams in Fig.~\ref{fig:disso}, for kinematics where the
energy-transfer and momentum-transfer to the hadronic system are
equal, i.e. light-like. We find
\begin{eqnarray}
X & = & 
2 i \sqrt{\pi\over\gamma^3} {1\over\sqrt{1-\gamma r_3}}
{1\over\left[ -{1\over a_1} + {1\over 2} r_1 |{\bf p}|^2 - i |{\bf p}|\right]}
\nonumber\\
&&
\left[\ 
\langle x^{n-1}\rangle_{ q_A^{(1)}}\ {\gamma^2\over |{\bf p}|^2+\gamma^2}
\left( \gamma - {1\over a_1} +  {1\over 2} r_1 |{\bf p}|^2 \right)
\ +\ {1\over 2}\gamma^2\sqrt{r_1 r_3} \
 \tilde\sigma^{(n)}_{q}
\ \right]
\ \ \ ,
\label{eq:dvcsX}
\end{eqnarray}
where ${\bf p}$ is the momentum of one of the nucleons in the final
state~\footnote{Notice that while this form-factor sums all orders in
$|{\bf p}|/\gamma$, it is perturbative in the expansion parameter
$|{\bf p}|/m_\pi$.}.  
While the spin-dependent structure functions determine the single-nucleon
contributions to the deuteron break-up (first term in the square brackets 
of eq.~(\ref{eq:dvcsX}) proportional to $\langle x^{n-1}\rangle_{ q_A^{(1)}}$
), there is an additional
contribution coming from short-distance strong interactions
(second term in the square brackets of eq.~(\ref{eq:dvcsX})
proportional to $\tilde\sigma^{(n)}_{q}$).
This amplitude is similar in form to that of $np\rightarrow
d\gamma$~\cite{Chen:1999bg,Rupak:1999rk} for obvious reasons~\footnote{
Both terms in the square brackets of eq.~(\ref{eq:dvcsX}) naively appear
to be of the same order in the expansion, however one power of the expansion parameter, $Q$,
has been hidden in $\tilde\sigma^{(n)}_{q}$ for ease of notation.}. 

We emphasize that the validity of $\nopit$ requires that the energy
imparted to the hadronic system be significantly less than $m_\pi$.
For higher energies, pions must be included explicitly in the
description.  Even with the inclusion of pions as dynamical fields,
local four-nucleon operators will continue to contribute to nuclear
matrix elements of the twist-2 operators. In the pionful theory these
operators describe contributions from physics at scales of the order of
and smaller than the chiral symmetry breaking scale.  It is important
to point out that the same two-body operators that are inserted in the
deuteron will contribute to DVCS on the triton and $^3$He, with
three-body operators expected to be suppressed by their naive
engineering dimensions~\cite{Bedaque:1998km,Bedaque:1999ve}.

In this work we have discussed the EMC effect in terms of the matrix
elements of local operators. We have made no attempt to determine the
underlying mechanism; rather we have demonstrated how to describe the
effect in a model-independent manner. As for the case of the single
nucleon, only positive moments of the parton distributions can be
related to the forward matrix elements of local operators, and so it
is natural to consider only the positive moments of the parton
distributions of nuclei, including the deuteron. By contrast, the
small-$x$ behavior is not related to forward matrix elements of local
operators, and arises from long-distance strong-interaction
contributions to the electromagnetic-nucleus interactions.  So while
the positive moments can be considered properties of nucleons and
nuclei, the small-$x$ behavior is a property of the interaction matrix
element, and in general, is different for each probe.  For forward
processes and those involving only small momentum transfers we have
used $\nopi$ and $\nopit$ to compute the EMC effect and the
DVCS-dissociation of the deuteron in terms of single-nucleon
contributions and four-nucleon interactions due to short-distance
physics.  We consider this paper to be the first step toward a
detailed calculation of the dissociation process with realistic kinematics.  We
estimate that a measurement of the break-up cross sections will show
deviations from the single-particle estimate at the level of
$5\%\rightarrow 10\%$, in analogy with the process $np\rightarrow
d\gamma$, and in keeping with the EMC effect.  Measurement of this
break-up cross section at Jefferson Laboratory or elsewhere, and a
determination of the deviation from the single-nucleon effect, would
provide a fascinating probe of the strong interactions.

\vfill\eject

\acknowledgments

\noindent 
We are grateful to Jerry Miller for comments and discussions which
have substantially increased the quality of the manuscript. We also
thank Bill Donnelly for valuable discussions. MJS would like to thank
the Center for Theoretical Physics at MIT for kind hospitality during
the completion of this work.  The work of MJS is supported in part by
the U.S.~Dept. of Energy under Grant No.~DE-FG03-97ER4014.  The work
of SRB is supported in part by the National Science Foundation under
grant No. PHY-0400231 and by DOE contract DE-AC05-84ER40150, under
which the Southeastern Universities Research Association (SURA)
operates the Thomas Jefferson National Accelerator Facility.


\end{document}